\documentclass[12pt]{article}
\usepackage[utf8]{inputenc}
\usepackage[left=3cm,right=3cm,top=3cm,bottom=3cm]{geometry}
\usepackage{lmodern}
\usepackage{tabularx}
\usepackage{soul}
\usepackage{makecell} 
\usepackage{graphicx}
\usepackage{tikz}
\usetikzlibrary{shapes.geometric, arrows.meta, positioning}
\usepackage[dvipsnames]{xcolor}
\usepackage{array}
\usepackage{dirtytalk}
\usepackage{adjustbox}
\usepackage{setspace} 
\usepackage{hhline} 
\usepackage{rotating}
\usepackage{float} 
\usepackage{array} 
\usepackage{eurosym} 
\usepackage{arydshln} 
\usepackage{longtable} 
\usepackage{csquotes} 
\usepackage{nccmath} 
\usepackage{bm} 
\usepackage{booktabs} 
\usepackage{xcolor} 
\usepackage{multirow} 
\usepackage{bm} 
\usepackage{physics} 
\usepackage{systeme} 
\usepackage{mathtools} 
\usepackage{amsmath, amsthm, amssymb, amsfonts} 
\usepackage{graphicx} 
\usepackage{amsmath}
\usepackage{amssymb}
\usepackage{ulem}

\usepackage{graphicx}
\usepackage{subcaption}
\usepackage[round, authoryear]{natbib}
\usepackage{ulem}
\usepackage{pgfplots} 
\pgfplotsset{compat=1.16} 
\usepackage{tikz} 
\usetikzlibrary{intersections,calc} 
\usepackage{caption} 
\usepackage{soul} 
\usepackage{sidecap} 
\usepackage[space]{grffile} 
\usepackage{subcaption} 
\usepackage{makebox} 
\usepackage[short]{optidef} 
\usepackage{abraces} 
\usepackage{enumitem} 
\usepackage{comment} 
\usepackage{cancel} 
\usepackage{accents} 
\usepackage{stackrel} 
\usepackage{dashrule} 
\usepackage{scalerel} 
\usepackage{centernot} 
\usepackage{pifont} 
\usepackage{pdflscape}
\usepackage{chngcntr}
\usepackage[flushleft]{threeparttable}
\usepackage[round]{natbib}
\usepackage{hyperref,etoolbox}
\usepackage{titlesec}

\usepackage{newtxtext,newtxmath} 

\usepackage[british]{babel}

\hypersetup{
    colorlinks=true,
    linkcolor=black,
    filecolor=magenta,
    citecolor=black, 
    urlcolor=black,
    }
\makeatletter
\def\@footnotecolor{red}
\define@key{Hyp}{footnotecolor}{%
 \HyColor@HyperrefColor{#1}\@footnotecolor%
}
\patchcmd{\@footnotemark}{\hyper@linkstart{link}}{\hyper@linkstart{footnote}}{}{}
\makeatother
\hypersetup{footnotecolor=black}

\usepackage[ruled,vlined]{algorithm2e}
\SetKwInput{KwInput}{Input}                
\SetKwInput{KwOutput}{Output}

\usetikzlibrary{arrows,decorations.markings}
\usetikzlibrary{decorations.pathreplacing}
\tikzset{myarr/.style={decoration={markings, mark=between positions 0 and 1 step 4mm with {\arrow{stealth}},},postaction=decorate}}
\tikzset{myarrrev/.style={decoration={markings, mark=between positions 0 and 1 step 4mm with {\arrowreversed{stealth}},},postaction=decorate}}
\DeclareCaptionLabelFormat{subfig}{\figurename #1~\thefigure.\alph{subfigure}:}

\theoremstyle{definition}

\theoremstyle{definition}

\theoremstyle{definition}

\theoremstyle{definition}

\theoremstyle{definition}

\theoremstyle{theorem}

\theoremstyle{definition}

\theoremstyle{definition}

\definecolor{blue}{RGB}{0,114,178}
\definecolor{red}{RGB}{204,51,17}
\definecolor{yellow}{RGB}{240,228,66}
\definecolor{green}{RGB}{0,158,115}

\tikzset{
solid node/.style={circle,draw,inner sep=1.5,fill=black},
hollow node/.style={circle,draw,inner sep=1.5}
}

\title{Income Inequality and Economic Growth: A Meta-Analytic Approach}

 \vspace{5pt}
\date{This version: \today}
 
\begin{document}

\setcounter{page}{0}
\pagenumbering{Alph}

\doublespacing
\begin{center}
\Large{\textbf{Income Inequality and Economic Growth: A Meta-Analytic Approach}}\\[.25cm]
\large{%
\begin{tabular}{ccc}
Lisa Capretti$^{a}$
& %
Lorenzo Tonni$^{b}$

\end{tabular}
}
\\
\vspace{1cm}
\normalsize{\today}\\
\vfill
\end{center}
\footnotesize{\textbf{Abstract.} The empirical literature on the relationship between income inequality and economic growth has produced highly heterogeneous and often conflicting results. This paper investigates the sources of this heterogeneity using a meta-analytic approach that systematically combines and analyzes evidence from relevant studies published between 1994 and 2025.
We find an economically small but statistically significant negative average effect of income inequality on subsequent economic growth, together with strong evidence of substantial heterogeneity and selective publication based on statistical significance, but no evidence of systematic directional bias.
To explain the observed heterogeneity, we estimate a meta-regression. The results indicate that both real-world characteristics and research design choices shape reported effect sizes. In particular, inequality measured net of taxes and transfers is associated with more negative growth effects, and the adverse impact of inequality is weaker -- or even reversed -- in high-income economies relative to developing countries. Methodological choices also matter: cross-sectional studies tend to report more negative estimates, while fixed-effects, instrumental-variable, and GMM estimators are associated with more positive estimates in panel settings.}
\vfill
\small{\noindent\textbf{Key Words:} Meta-Analysis; Inequality; Growth; Publication-Bias.}\\\vspace{\fill}

\footnotesize{%
\noindent $^{(a)}$ CEIS, University of Rome Tor Vergata, Italy. Email: lisa.capretti@uniroma2.it\\
\noindent $^{(b)}$ Department of Economics, Management and Quantitative Methods, University of Milan, Italy. Email: lorenzo.tonni@unimi.it (corresponding author) \\

\vfill

\thispagestyle{empty}
\newpage
\setcounter{footnote}{0}
\pagenumbering{arabic}
\doublespacing
\normalsize

\newpage


\section{Introduction}

In recent decades, the unprecedented availability of income distribution data has fueled a surge in studies examining the role of income inequality in economic growth. 
This body of literature is characterized by substantial heterogeneity in results and has not converged to a clear consensus regarding either the sign or the magnitude of the effect of income inequality on subsequent economic growth. This heterogeneity may reflect differences in econometric specifications, data sources, and country samples, as well as the existence of multiple underlying mechanisms through which income distribution affects growth, whose relative importance may vary across countries. This paper aims to shed light on the factors behind this heterogeneity by means of meta-analytic tools, which allow for the systematic combination and quantitative analysis of results from the existing empirical studies.

The theoretical literature has identified several channels through which income inequality can influence economic growth. Most of these channels imply a negative impact of the former on the latter. Prominent examples include credit constraints \citep{galor1993income, banerjee1993occupational, piketty1997dynamics, galor2004physical}, fiscal policy \citep{persson1991inequality}, fertility rates \citep{de2003inequality}, domestic demand \citep{murphy1989income}, social capital \citep{knack1997does}, crime rates \citep{josten2003inequality}, social instability \citep{alesina1996income, knack1997does}, and corruption \citep{esteban2006inequality, galor2009inequality}. In contrast, other contributions suggest that income inequality may foster growth. For instance, bottom-up redistribution could stimulate capital accumulation if the savings function is convex \citep{bourguignon1981pareto, galor2004physical}, or provide greater incentives for research and development \citep{foellmi2006income}. More generally, income distribution could serve as an incentive for the optimal allocation of productive factors, resulting in an equity-efficiency trade-off \citep{mirrlees1971exploration, lazear1981rank, rebelo1991long, okun2010equality}.

The task of identifying the sign and magnitude of the net effect of income inequality on economic growth has fallen to the empirical analysis. The literature has developed along two main lines.\footnote{For a detailed discussion see \citet{baselgia2023inequality}.} The first strand focuses on single transmission channels and tipically employs two-stage structural models, in which a mediator variable is first regressed to income inequality and subsequently linked to economic growth. 
The second strand employs a reduced-form approach, estimating the overall effect of income inequality on growth by regressing GDP per capita growth on an inequality index and a set of control variables.
This paper exclusively focuses on the reduced-form strand for two key reasons. Firstly, results from the two approaches are not directly comparable. Secondly, the number of studies relying on two-stage structural models is too limited to conduct a separate meta-analysis.

Meta-analysis provides a natural framework to reconcile the diverging findings in the inequality–growth literature. By systematically collecting results from studies that meet explicitly stated inclusion criteria and applying standardized statistical procedures, it reduces the scope for selective interpretation and allows a robust assessment of the differences in reported results \citep{stanley2012meta}. Given the pronounced variability that characterizes empirical estimates of the inequality–growth relationship, a meta-analytic approach is particularly well suited to identify the driving factors of this heterogeneity.

Previous meta-analyses provide conflicting evidence regarding the sources of heterogeneity in the inequality-growth relationship. While \citet{de2008meta} identify the econometric estimator -- and specifically fixed effects -- as the main driver of variation, \citet{neves2016meta} attribute heterogeneity primarily to the dataset structure (panel vs. cross-section) rather than the estimator employed. Furthermore, the two studies disagree on the role of data quality. However, they agree that income inequality is more detrimental to growth in developing countries and both document the presence of publication bias, albeit in different forms: selective publication based on statistical significance in \citet{neves2016meta} and a bias toward negative estimates in \citet{de2008meta}. Given these inconsistencies, we contribute to this debate by providing an updated and methodologically refined meta-analytic assessment.
 In particular, our analysis improves upon previous works in several ways. First, we update the pool of observations to include studies published up to 2025, thereby incorporating more than a decade of recent research. Compared to \citet{neves2016meta}, our dataset adds 11 years of literature, while relative to \citet{de2008meta} it adds 19 years.\footnote{The larger number of studies in \citet{de2008meta} reflects the inclusion of working papers, whereas we focus exclusively on peer-reviewed journal articles. See Section~\ref{Meta_Dataset} for details.} Second, unlike \citet{neves2016meta}, we collect all reported estimates from each study rather than relying on author-selected preferred specifications. This results in a substantially larger dataset of 531 estimates (compared to 49), increasing statistical power and reducing the influence of individual studies. Third, in contrast to \citet{neves2016meta}, we restrict attention to estimates based exclusively on the Gini coefficient, thereby ensuring both conceptual and quantitative comparability across studies. Different inequality measures capture distinct aspects of the income distribution and are not simple rescalings of one another. Combining estimates based on heterogeneous inequality indices implicitly assumes that these measures are interchangeable, an assumption that is difficult to justify and may confound true economic heterogeneity with measurement differences. Fourth, we employ state-of-the-art methods to address publication bias, relying on the RoBMA–PSMA framework, and conduct a meta-regression analysis that includes a broader set of moderators.

Our results indicate that the average effect of income inequality on economic growth is negative and statistically significant, although economically small. We also find substantial between-study heterogeneity, consistent with the presence of multiple true effect sizes. The publication-bias analysis reveals strong evidence of selection based on statistical significance, pointing to an over-representation of significant results in the literature, but no evidence of systematic directional bias. To investigate the sources of heterogeneity, we estimate multilevel meta-regressions that account for dependence among multiple estimates reported within the same study. These results show that both real-world characteristics and research design choices contribute to variation in reported effects. In particular, inequality measured net of taxes and transfers is associated with more negative growth effects, and the adverse impact of inequality is stronger in developing countries. Methodological choices also matter: cross-sectional studies tend to report more negative estimates, an effect that is mitigated by the inclusion of regional controls, while Instrumental Variable (IV), fixed-effects, and Generalized Method of Moments (GMM) estimators are associated with more positive estimates in panel settings. By contrast, growth horizon length, data quality, and journal characteristics do not appear to systematically influence reported results.

The remainder of the paper is organized as follows. Section~\ref{empirical_literature} reviews the empirical literature estimating the effect of income inequality on economic growth. Section~\ref{Meta_Dataset} describes the study selection process and presents descriptive statistics summarizing the characteristics of the meta-analytic dataset. Section~\ref{meta_analysis} reports the meta-analysis. We first estimate the average effect size (Section~\ref{preliminary_analysis}), then test for the presence of publication bias (Section~\ref{publication_bias}), and finally conduct a meta-regression to investigate the sources of heterogeneity across studies (Section~\ref{meta_regression}). Section~\ref{conclusion} concludes.

\section{Empirical evidence}
\label{empirical_literature}

The first empirical studies investigating the relationship between income inequality and economic growth emerged in the early 1990s, following the release of the first large-scale datasets on income distribution. This early strand of the literature \citep{alesina1994distributive, persson1991inequality, perotti1996growth, clarke1995more} relies on ad hoc datasets constructed by aggregating inequality measures from multiple sources. Due to limited time coverage, these studies primarily employ cross-sectional models. Typically, the growth rate of GDP per capita is regressed on an inequality index and a set of control variables, as illustrated by the following specification:

\begin{equation}
\label{model_cross_section}
    y_i = \alpha + \sum_{j=1}^{J} \theta_j I_{ji} + \sum_{k=1}^{K} \beta_{k} X_{ki} + \varepsilon_i
\end{equation}

where $\alpha$ is a constant term, $y_i$ denotes the growth rate of GDP per capita in country $i$, typically measured over a long horizon of 25–40 years, and $I_{ji}$ represents inequality index $j$ for country $i$, usually observed at the beginning of the sample period to mitigate concerns of reverse causality. The associated coefficient $\theta_j$ captures the effect of inequality on growth. Although the Gini coefficient is the most commonly used measure, several studies also rely on income shares, decile ratios, and other summary indicators of inequality. The vector $X_{ki}$ includes standard growth controls such as initial income, education, and investment. Overall, this early literature predominantly finds that higher income inequality is associated with lower subsequent economic growth.

This negative relationship was challenged in the late 1990s with the emergence of studies based on panel data. The release of the \citet{deininger1996new} dataset provided repeated observations of income distribution over time, enabling researchers to exploit within-country variation. This development gave rise to panel regressions in which the sample period is divided into shorter subperiods - typically five to ten years - rather than a single long growth spell. Studies in this second wave of the literature \citep{li1998income, partridge1997inequality, forbes2000reassessment}, often employing fixed-effects estimators, tend to find a positive effect of income inequality on subsequent economic growth.

Subsequent research, however, has questioned the robustness and interpretation of these findings. Some contributions argue that the positive relationship uncovered in early panel studies is spurious. \citet{banerjee2003inequality} depart from the linear framework in equation~(\ref{model_cross_section}) and document a non-linear relationship, whereby any change in inequality - regardless of its direction - reduces subsequent growth. They attribute this pattern to measurement error in inequality, which tends to be more severe during periods of economic distress characterized by sharp output declines. Similarly, \citet{scholl2019re} challenge the foundations of the early panel evidence, showing that the positive association between inequality and growth is largely driven by transition economies in Eastern Europe during the 1990s. In their interpretation, the collapse of the Soviet Union led simultaneously to rising inequality and subsequent economic recovery, generating a correlation that does not reflect a causal relationship.

Other studies suggest that the relationship between inequality and growth is inherently context-dependent. Several authors find that once countries are grouped by their level of development, inequality promotes growth only in high-income economies, while it hampers growth in less developed ones \citep{barro2000inequality, khalifa2010income, grundler2018growth}. In contrast, \citet{brueckner2018inequality} report the opposite pattern, with inequality fostering growth in low-income countries but reducing it in high-income economies.

The impact of inequality may also depend on which segment of the income distribution is considered. \citet{voitchovsky2005does} argues that relying on a single inequality index masks important distributional dynamics and shows that inequality at the top of the distribution can stimulate growth, whereas inequality at the bottom is detrimental. Related evidence is provided by \citet{bartak2020inequality}, who document heterogeneous effects across different parts of the income distribution, although their overall results point to a negative impact of inequality on growth. Along similar lines, \citet{marrero2022growth} find that inequality has no direct effect on growth but operates indirectly through its association with poverty, which significantly reduces economic development.

Finally, the time horizon over which growth is measured appears to matter. \citet{herzer2012inequality} criticize the common practice of regressing growth rates on inequality levels and instead apply cointegration techniques using variables in levels. Their results indicate a negative long-run relationship between inequality and economic growth. Comparable conclusions are reached by \citet{berg2018redistribution}, who identify a negative long-term effect and a positive short-term effect of inequality. These findings are broadly consistent with those of \citet{halter2014inequality} and \citet{davis2011institutional}. Evidence in \citet{el2019impact} further suggests that inequality may stimulate growth over intermediate horizons.

\section{Meta-Dataset}
\label{Meta_Dataset}

This section describes the construction of our meta-analytic dataset and summarizes its main characteristics. We searched for English-language peer-reviewed articles indexed in the Scopus database using the keyword \textit{``growth''} in combination with either \textit{``inequality''} or \textit{``distribution''} in titles or abstracts. We complemented this search with a title-based query on Google Scholar. Overall, this procedure yielded 3,461 records (3,245 from Scopus and 216 from Google Scholar). We additionally screened references cited in the selected articles and included all studies analyzed by \citet{de2008meta} and \citet{neves2016meta} that were not already part of our sample and satisfied our inclusion criteria -- described below in detail -- adding 25 further studies.

After screening abstracts, we excluded 3,438 articles that did not examine the relationship between income inequality and economic growth. The full texts of the remaining 48 studies were then assessed in detail. We excluded studies and estimates that did not conform to the reduced-form framework described in equation~(\ref{model_cross_section}) or its panel-data counterpart.\footnote{While acknowledging their contribution to the literature, studies estimating non-linear relationships were excluded because their results are not directly comparable with the bulk of the empirical literature and their number is insufficient to support a separate meta-analysis.} We also excluded studies conducted exclusively at the national level to avoid confounding effects driven by country-specific institutional or regional factors. Furthermore, studies measuring inequality using concepts other than income - such as wealth, land, or human capital - were not considered.\footnote{The limited number of such studies does not allow us to control for those specific concepts of inequality in the meta-analysis.}

To ensure comparability across studies, all included estimates were required to meet a set of minimum specification criteria. First, we retained only estimates based on the Gini coefficient.\footnote{Alternative inequality measures, such as the Theil index or decile ratios, capture different features of the income distribution. Pooling estimates based on heterogeneous inequality measures would therefore confound substantive heterogeneity with measurement differences. For comparison purposes with \citet{neves2016meta}, we replicate the meta-regression analysis including non-Gini-based estimates in Appendix.} Second, we excluded estimates that did not control for initial GDP per capita, given its well-known role in the growth dynamics since the seminal contribution of \citet{solow1956contribution}. Finally, we removed outliers to prevent extreme values from disproportionately influencing the results. An estimate was classified as an outlier if either its effect size or its standard error deviated from the sample median by more than ten interquartile ranges \citep{mccracken2016fred}. Applying these minimum criteria led to the exclusion of an additional 15 studies. Figure~\ref{fig:prisma} in Appendix~\ref{app::prisma} summarizes the search and screening process using a PRISMA flow diagram.

The final dataset comprises 33 published articles and a total of $N = 531$ estimates. The effect size extracted from each study is the coefficient $\theta_j$ associated with the Gini index $I_{ji}$ in equation~(\ref{model_cross_section}). This coefficient measures the change in the annual growth rate of GDP per capita associated with a one-point increase in the Gini index, measured on a 0–100 scale.\footnote{For example, a coefficient of $\theta = -0.0345$ implies that a one-point increase in the Gini index reduces annual GDP per capita growth by 0.0345 percentage points. When studies employ alternative scaling conventions, estimates are rescaled to ensure consistency across observations.}

Table~\ref{tab::dataset} reports key characteristics of the studies included in the meta-analysis, including the number of estimates per study (Observations, Column~3), the structure of the dataset (cross-sectional or panel, Column~4), the average effect size within each study $\theta$ (Column~5), whether inequality is measured before or after taxes and transfers (Income Concept, Column~6), the type of countries included (Column~7), and the estimation methods employed (Column~8). Additional information on the collected variables and their summary statistics is provided in Appendix \ref{app::descriptives}.\footnote{All authors jointly conducted the literature search, independently reviewed the included studies, and collaborated in coding the extracted estimates. Coding decisions were cross-validated to ensure accuracy and consistency.}

\newgeometry{left=1.5cm,right=1.5cm,top=3cm,bottom=3cm}

\begin{table}[h]
\small
\caption{Dataset characteristics}
\begin{threeparttable}
\resizebox{7.4in}{!}{%
\begin{tabular}{ | c | l | c | l | c | c | l | l | }
\hline
\hline
\textbf{N} & \textbf{Study} & \textbf{Observations} & \textbf{Structure} & \textbf{Average $\theta$} & \textbf{Income Concept} & \textbf{Country type} & \textbf{Estimation Method} \\
\hline
\hline
1&Clarke (1995)&7&CS&-.076&Gross, Net&2&OLS,IV,WLS \tabularnewline
2&Galor and Zang (1997)&14&CS&-.051&Gross&1&OLS \tabularnewline
3&Li and Zou (1998)&40&CS, Panel&.087&Gross, Net&2&OLS,FE,RE \tabularnewline
4&Deininger and Squire (1998)&12&CS&-.024&Gross, Net&1&OLS \tabularnewline
5&Tanninen (1999)&5&CS&-.138&Gross, Net&2&OLS \tabularnewline
6&Knell (1999)&3&CS&-.044&Unspecified&1&OLS \tabularnewline
7&Mo (2000)&20&CS&-.195&Unspecified&2&IV \tabularnewline
8&Forbes (2000)&50&CS, Panel&.150&Unspecified&1&OLS,FE,RE,GMM \tabularnewline
9&Sylwester (2000)&6&CS&-.053&Gross, Net&2&OLS,IV \tabularnewline
10&Barro (2000)&6&Panel&-.013&Gross&2&RE,IV \tabularnewline
11&Keefer and Knack (2002)&2&CS&-.067&Unspecified&2&OLS \tabularnewline
12&Castelló  and Doménech (2002)&3&CS&.306&Unspecified&2&OLS \tabularnewline
13&Banerjee and Duflo (2003)&14&Panel&.112&Unspecified&2&FE,RE,GMM \tabularnewline
14&Odedokun and  Round (2004)&2&CS&-.012&Unspecified&2&OLS \tabularnewline
15&Bleaney and Nishiyama (2004)&16&CS&.016&Gross&2&OLS \tabularnewline
16&Voitchosky (2005)&14&Panel&-.040&Net&1&GMM \tabularnewline
17&Knowles (2005)&6&CS&-.057&Gross, Net&2&OLS \tabularnewline
18&Sakar (2007)&2&CS&-.056&Unspecified&2&OLS \tabularnewline
19&Noh and Yoo (2008)&4&Panel&.277&Unspecified&2&FE \tabularnewline
20&Huang, Lin and Yeh (2009)&4&Panel&.062&Unspecified&1&OLS,SEM-GMM \tabularnewline
21&Castellò-Climent (2010)&42&Panel&-.003&Unspecified&1&GMM \tabularnewline
22&Chambers and Krause (2010)&3&Panel&.002&Gross, Net&2&GMM,\tabularnewline
23&Davis and Hopkins (2011)&25&CS, Panel&-.006&Unspecified&2&OLS,FE,,RE,BE,SEM-OLS \tabularnewline
24&Woo (2011)&18&CS&-.067&Unspecified&2&OLS \tabularnewline
25&Herzer and Vollmer (2012)&6&Panel&-.012&Gross&1&ECM \tabularnewline
26&Halter et al. (2014)&32&Panel&-.006&Gross&2&GMM \tabularnewline
27&Brueckner and Lederman (2018)&6&Panel&.233&Unspecified&1&FE,IV \tabularnewline
28&Grundler and Scheuermayer (2018)&34&Panel&-.157&Net&2&GMM \tabularnewline
29&Ostry et al. (2018)&22&Panel&-.152&Net&2&GMM \tabularnewline
30&Bartak and Jabłoński (2018)&6&Panel&-.172&Net&2&GMM \tabularnewline
31&Scholl and Klasen (2019)&8&Panel&-.075&Net&2&FE,IV,GMM \tabularnewline
32&El-Shagi and Shao (2019)&2&Panel&.699&Unspecified&2&FE \tabularnewline
33&Marrero and Serven (2021)&97&Panel&-.030&Net&1&OLS,FE,GMM \tabularnewline
\hline
\end{tabular}%
}
\begin{tablenotes}
\scriptsize
\item \begin{minipage}[t]{1\textwidth}
Notes: In Column 3 (Structure): CS stands for Cross-Sectional. In Column 7 (Country type): (1) refers to the inclusion of both high-income and least developed countries, controlling for the level of development; 
(2) to the inclusion of both types of countries without controlling for the level of development; 
In Column 8 (Estimation Method): WLS stands for Weighted Least Squares; 
FE for Fixed Effects; RE for Random Effects; BE for Between Estimation; ECM for Error Correction Model.
\end{minipage}
\end{tablenotes}
\end{threeparttable}
\label{tab::dataset}
\end{table}

\newgeometry{left=3cm,right=3cm,top=3cm,bottom=3cm}

The number of studies included in our dataset differs from \citet{de2008meta} and \citet{neves2016meta}, even when the period of analysis overlaps. In particular, some studies included in \citet{de2008meta} do not appear in our dataset since they are unpublished working papers instead of peer-reviewed journal articles. We restrict attention to published studies in order to ensure minimum standards of methodological scrutiny concerning data quality, econometric set-up and robustness transparency. Additionally we exclude, some studies analyzed by \citet{neves2016meta} as they rely on inequality indices other than the Gini index.

Figure~\ref{fig:forest_plot} displays the distribution of the collected effect sizes, sorted by magnitude. The estimates span both positive and negative values and cluster around zero, offering no clear indication of the sign or magnitude of the average effect. This dispersion further motivates the meta-analytic and meta-regression analyses presented in the next section.

\begin{figure}[H]
\centering
\setlength{\tabcolsep}{.008\textwidth}

\includegraphics[scale=0.6]{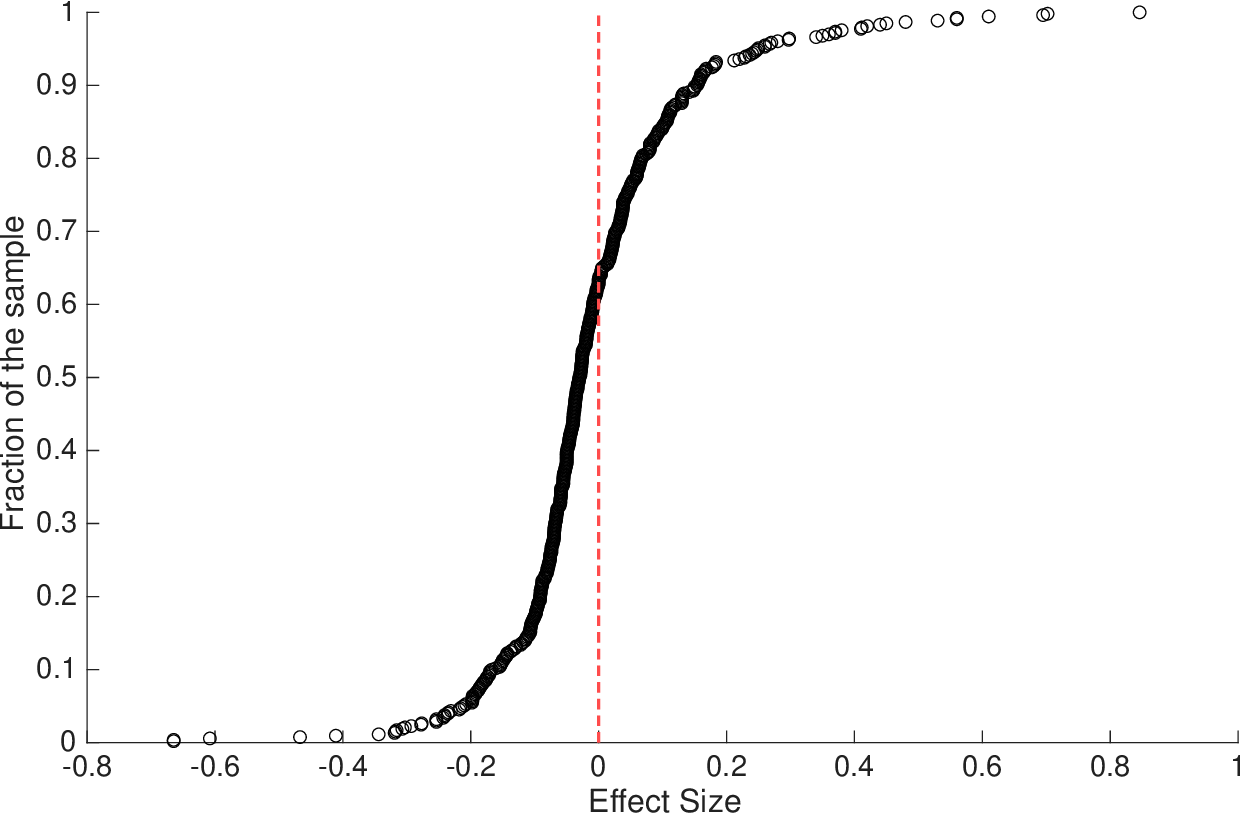}

\caption{Forest Plot}
\label{fig:forest_plot}
\end{figure}

\section{Meta-Analysis}
\label{meta_analysis}

\subsection{Preliminary analysis}
\label{preliminary_analysis}

A central objective of a meta-analysis is to estimate the average effect size implied by the empirical literature. Let $\theta_q$ denote the $q$th reported effect size and let $w_q$ be a precision weight. The precision-weighted mean is given by

\begin{equation}
\label{eq:overall_effect}
    \hat{\mu} = \frac{\sum_{q=1}^{N} w_q \theta_q}{\sum_{q=1}^{N} w_q},
\end{equation}

where $\hat{\mu}$ is the overall size effect, and $w_q$ is the associated weight given by the inverse of its variance, $1/\sigma^2_q$.

We estimate $\mu$ using the Unrestricted Weighted Least Squares (UWLS) approach of \citet{stanley2015neither}. UWLS treats each reported estimate as an observation of an underlying average effect and allows for heteroscedasticity proportional to the reported sampling variances. Operationally, UWLS is implemented as a weighted regression of $\theta_q$ on a constant, using weights proportional to $1/\sigma_q^2$, and inference is based on study-level cluster-robust standard errors to account for multiple, non-independent estimates reported within the same study.

The UWLS results yield $\hat{\mu} = -0.019$ (p-value $<0.01$), implying that a one-point increase in the Gini index (on a 0--100 scale) is associated, on average, with a 0.019 percentage-point reduction in the annual growth rate of GDP per capita. Although statistically significant, the estimated average effect is economically small.

\subsection{Publication-bias}\label{publication_bias}

Publication bias arises when the probability that a result appears in the published literature depends on its statistical significance or on how well it conforms to prevailing expectations. Such selectivity can distort the evidence base and bias estimates of the average effect. A common diagnostic is the \textit{small-study effect}: estimates from less precise studies (with larger standard errors) are expected to be more dispersed around the underlying effect. In the absence of selective publication, the distribution of effect sizes should be approximately symmetric around the mean across the range of standard errors. If, however, statistically insignificant or undesired results are less likely to be published, the observed distribution may become asymmetric.

Figure~\ref{fig:funnel_plot} presents a funnel plot of the collected estimates, plotting each effect size against its standard error. Visual inspection suggests some asymmetry, with a concentration of estimates in the statistically significant negative region, although the pattern is not sufficiently sharp to draw firm conclusions from the plot alone.

\begin{figure}
\centering
\caption{Funnel Plot}
\includegraphics[scale=0.6]{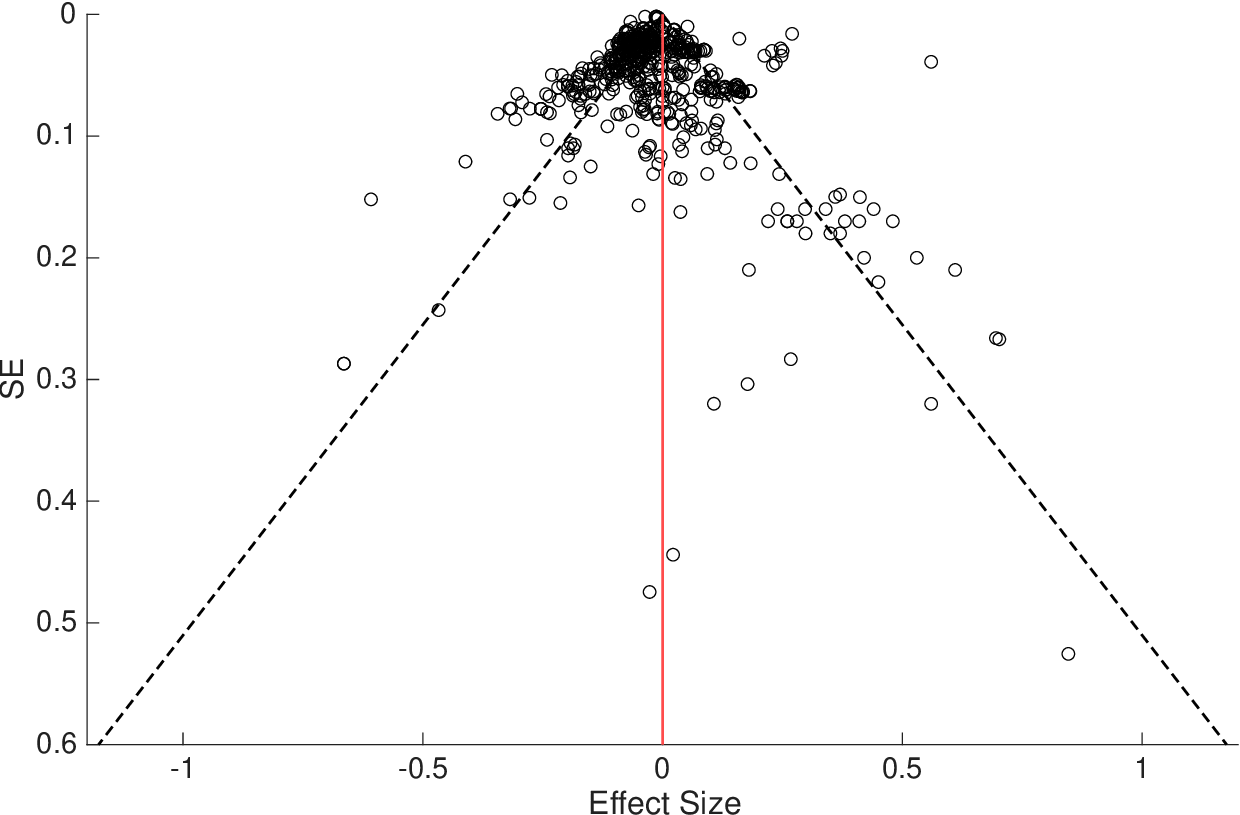}
\label{fig:funnel_plot}
\caption*{\scriptsize\textit{Notes}: dotted lines are the pseudo-95\% confidence intervals. SE stands for Standard Error.}
\end{figure}

Since funnel-plot interpretation is inherently subjective and small-study patterns may also arise from heterogeneity, we implement a formal publication-bias assessment using the Robust Bayesian Meta-Analysis Publication-bias and Small-study effects Model Averaging (RoBMA--PSMA) framework, following \citet{bartovs2023robust}.\footnote{We restrict attention to two-sided selection models because there is no strong a priori basis for assuming one-sided selection in a specific direction in this context.} RoBMA--PSMA combines multiple candidate meta-analytic models that differ in their assumptions about the presence of the effect, heterogeneity and selective publication and averages across them using posterior model probabilities. In particular, the model set includes standard meta-analytic specifications, selection models based on two-sided $p$-value intervals (weight functions), and regression-based adjustments for small-study effects (Precision-Effect Test, PET and Precision-Effect Estimate with Standard Errors, PEESE). Further implementation details are reported in Appendix \ref{app:robma}.

\begin{table}[ht]
\centering
\caption{RoBMA-PSMA Analysis}
\label{tab:robma_summary}
\begin{threeparttable}
\footnotesize
\begin{tabular}{lcc}
\hline
\multicolumn{3}{l}{\textbf{Component summary}} \\
\hline
Component & Posterior probability & BF \\
\hline
Effect ($\mu \neq 0$)       & 0.999 & $1.45 \times 10^{3}$ \\
Heterogeneity ($\tau > 0$) & 1.000 & $3.27 \times 10^{240}$ \\
Bias          & 1.000 & $5.81 \times 10^{8}$ \\
\hline
\multicolumn{3}{l}{\textbf{Model-averaged estimates}} \\
\hline
Parameters & Mean & 95\% Credible Interval \\
\hline
$\hat{\mu}$   & $-0.016$ & $[-0.023,\,-0.010]$ \\
$\hat{\tau}$  & $0.066$  & $[0.059,\,0.074]$ \\
$\omega_{[0,0.05]}$   & $1.000$ & $[1.000,\,1.000]$ \\
$\omega_{[0.05,0.10]}$ & $0.739$ & $[0.512,\,0.962]$ \\
$\omega_{[0.10,1]}$    & $0.415$ & $[0.318,\,0.533]$ \\
PET    & $0.000$ & $[0.000,\,0.000]$ \\
PEESE  & $0.000$ & $[0.000,\,0.000]$ \\
\hline
\end{tabular}

\begin{tablenotes}
\item \textit{Notes:} Results are based on Bayesian model averaging over 20 candidate models obtained by combining the presence or absence of the overall effect ($\hat{\mu}$), heterogeneity ($\hat{\tau}$), and alternative publication-bias specifications (two-sided selection models based on the publication weight function $\omega$, PET, and PEESE).

\end{tablenotes}
\end{threeparttable}
\end{table}

 Table~\ref{tab:robma_summary} summarizes the RoBMA--PSMA results. They provide overwhelming posterior support for a non-zero overall effect, substantial heterogeneity, and selective publication. The Bayes Factors (BF) in Table~\ref{tab:robma_summary} (upper panel) imply that models allowing for heterogeneity and publication bias fit the observed distribution of effect sizes substantially better than models that omit these components.\footnote{They are largely above the conventional threshold of $BF>10$ \citep{bartovs2023robust}.}

The model-averaged estimate of the overall effect $\hat{\mu}$ is negative and close in magnitude to the UWLS estimate, indicating that the average effect, albeit significant, remains economically modest even after accounting for publication bias and heterogeneity. At the same time, the strong evidence for heterogeneity ($\hat{\tau}> 0$) suggests that a substantial share of the variation in $\theta_q$ reflects differences in underlying true effects across studies rather than sampling error alone.

The estimated publication weight function $\omega$ (also reported in Figure~\ref{fig:function_omega}) summarizes relative publication probabilities across two-sided $p$-value intervals. Estimates with $p$-values below 0.05 receive the highest relative publication weight, whereas less statistically significant results are assigned substantially lower weights. This pattern is consistent with selective publication, whereby authors or journals favor statistically significant findings irrespective of their sign, leading to an over-representation of statistically significant results in the published literature. This finding is in line with \citet{neves2016meta}.

\begin{figure}
\centering
\caption{Publication weight function}
\includegraphics[scale=0.6]{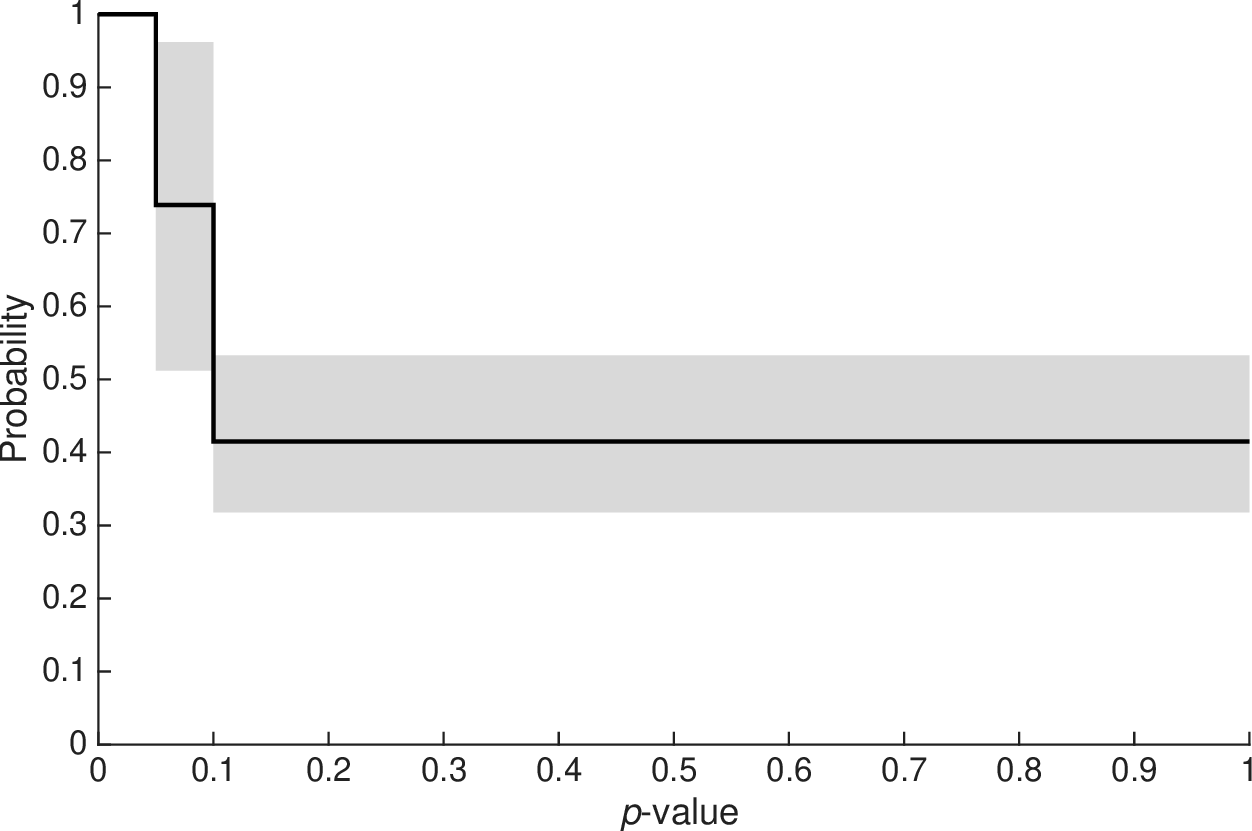}
\label{fig:function_omega}
\caption*{\scriptsize\textit{Notes}: Black line indicates the model-averaged weight function, grey area indicates the 95\% CI.}
\end{figure}

Finally, the regression-based small-study adjustments (PET and PEESE in Table~\ref{tab:robma_summary}) receive negligible posterior support, suggesting that precision-related small-study effects play a limited role in explaining the observed data. Instead, selective publication based on statistical significance provides a more plausible description of the publication process in this literature. This finding aligns with \citet{neves2016meta} and contrasts with \citet{de2008meta}, who report evidence consistent with a left-skewed small-study effect using PET-type tests.

The strong evidence for heterogeneity implies that the average effect, although statistically significant, is likely to mask substantial variation in true effects across studies. Explaining this heterogeneity is therefore the focus of the meta-regression analysis in the next section.

\subsection{Meta-Regression}
\label{meta_regression}

Meta-regression is a useful tool that aims to explain heterogeneity in reported effect sizes by relating them to observable study characteristics, commonly referred to as moderators. Formally, meta-regression is analogous to a standard regression model in which the dependent variable is the reported effect size and the covariates are study-level or estimate-level characteristics.

Meta-regression can be estimated under either a Fixed-Effects (FE) or a Random-Effects (RE) framework. The FE approach assumes that all heterogeneity in reported effects can be fully explained by the included moderators \citep{greenland1987quantitative}. In contrast, the RE approach allows for residual heterogeneity that remains unexplained by the observed moderators \citep{berkey1995random}. Given the complexity of the inequality--growth relationship and the unlikelihood that the available moderators exhaust all relevant sources of heterogeneity, we adopt a random-effects specification.

Let $\theta_q$ denote the $q$th reported effect size. The random-effects meta-regression can be written as
\begin{equation}
\theta_q = \beta_0 + \mathbf{M}_q \boldsymbol{\beta} + \xi_q,
\end{equation}
where $\beta_0$ is the conditional mean, $\mathbf{M}_q$ is a vector of moderators, $\boldsymbol{\beta}$ is the associated vector of coefficients, and $\xi_q$ is an error term with variance $\sigma_q^2 + \tau^2$, combining sampling uncertainty and unexplained heterogeneity. Estimation is performed using inverse-variance weights.

This formulation, however, treats all effect sizes as independent observations. In our dataset, individual studies report multiple estimates, implying that effect sizes are clustered within studies. Ignoring this dependence leads to over-representation of studies reporting many estimates and may bias inference. A common solution is to retain a single, representative estimate per study, as in \citet{neves2016meta}. While this approach avoids dependence, it comes at the cost of discarding a substantial amount of variation contained in the remaining estimates.

To address this issue, we adopt a multilevel (hierarchical) meta-regression model that explicitly accounts for the nesting of effect sizes within studies \citep{goldstein2011multilevel}. Let $\theta_{qr}$ denote the $q$th estimate reported in study $r$, with known sampling variance $\sigma_{qr}^2$. The model is specified as

\begin{equation}
\theta_{qr} = \beta_0 + \mathbf{M}_{qr}\boldsymbol{\beta} + u_r + w_{qr} + \varepsilon_{qr},
\end{equation}

where $u_r$ captures unobserved between-study heterogeneity, assumed to be normally distributed with variance $\tau^2_{\text{between}}$, and $w_{qr}$ captures within-study heterogeneity across multiple estimates from the same study, with variance $\tau^2_{\text{within}}$. The sampling error $\varepsilon_{qr}$ has mean zero and variance $\sigma_{qr}^2$. The model is estimated by Restricted Maximum Likelihood (REML). This three-level specification decomposes total variability in reported effects into sampling uncertainty, within-study heterogeneity, and between-study heterogeneity.

Given the large number of potential moderators and the risk of multicollinearity, we use Bayesian Model Averaging (BMA) as a data-driven filtering approach to select a parsimonious set of moderators \citep{steel2020model}. Specifically, we estimate a BMA regression with the reported effect size as the dependent variable and a broad set of candidate moderators as potential predictors. BMA evaluates many alternative model specifications and summarizes results by averaging across models, weighted by their posterior support in the data. We retain moderators with Posterior Inclusion Probabilities (PIP) above 0.1 and exclude the remaining candidates from the subsequent multilevel meta-regression. The full list of candidate moderators and their PIPs is reported in Appendix \ref{app:bma}. Table~\ref{tab:meta_reg} reports the included moderators along with their estimated meta-regression coefficients.

The selected moderators include variables that characterize four key features of the included studies: the characteristics of the dataset, the estimator employed, the covariates incorporated in the regression, and the statistics of the journal in which the study was published. Additionally, in order to control for potential direction-bias in terms of small-study effect, we included the standard error among the set of moderators.

\paragraph{Dataset characteristics}

Several features of the underlying dataset can shape the estimated inequality--growth relationship. First, the structure of the dataset matters. Cross-sectional designs are more exposed to omitted-variable bias and cannot control for time-invariant omitted variables as panel estimators. We therefore include a dummy variable $Cross-sectional$ equal to one for cross-sectional estimates and zero for panel-based estimates. The estimated coefficient is negative and highly significant, implying that cross-sectional specifications tend to report more negative effects of inequality on growth. This result is consistent with \citet{neves2016meta} but differs from \citet{de2008meta}, who find no systematic role for dataset structure.

\begin{table}[]
\centering
\caption{Multilevel Meta-Regression}
\label{tab:meta_reg}
\begin{threeparttable}
\footnotesize
\begin{tabular}{lc}
\toprule
Parameters & Coefficients \\
\midrule

\hspace{8mm}\textit{Standard error} 
& $0.0437$ \\
& $(0.1331)$ \\

\midrule

\textbf{Dataset characteristics} & \\

\hspace{8mm}\textit{Cross-sectional} 
& $-0.0674^{***}$ \\
& $(0.0168)$ \\

\hspace{8mm}\textit{Growth Span} 
& $-0.00027$ \\
& $(0.00074)$ \\

\hspace{8mm}\textit{High-income} 
& $0.0239^{***}$ \\
& $(0.0089)$ \\

\hspace{8mm}\textit{Net Inequality} 
& $-0.0597^{**}$ \\
& $(0.0261)$ \\

\hspace{8mm}\textit{High-quality data} 
& $-0.0216$ \\
& $(0.0337)$ \\

\hspace{8mm}\textit{Number of years} 
& $0.00040$ \\
& $(0.00073)$ \\

\midrule
\textbf{Estimator} & \\

\hspace{8mm}\textit{IV $\times$ cross} 
& $-0.0376$ \\
& $(0.0382)$ \\

\hspace{8mm}\textit{IV $\times$ panel} 
& $0.0803^{*}$ \\
& $(0.0446)$ \\

\hspace{8mm}\textit{Difference GMM} 
& $0.1312^{***}$ \\
& $(0.0154)$ \\

\hspace{8mm}\textit{Fixed Effects} 
& $0.1265^{***}$ \\
& $(0.0112)$ \\

\midrule
\textbf{Covariates} & \\

\hspace{8mm}\textit{Regional dummy $\times$ cross} 
& $0.0407^{*}$ \\
& $(0.0210)$ \\

\midrule
\textbf{Journal characteristics} & \\

\hspace{8mm}\textit{Simple Impact Factor (2022)} 
& $0.00655$ \\
& $(0.00640)$ \\

\midrule
Observations & 531 \\
Studies & 33 \\
\bottomrule
\end{tabular}

\begin{tablenotes}
\item \textit{Notes:} Standard errors in parentheses. 
$^{*} p<0.10$, $^{**} p<0.05$, $^{***} p<0.01$. 
Results are based on a multilevel REML meta-regression with random intercepts at the study and estimate levels.
\end{tablenotes}
\end{threeparttable}
\end{table}

Second, the time horizon over which growth is measured may affect results, as some channels may operate over different horizons \citep{herzer2012inequality, berg2018redistribution}. We therefore include the variable \textit{Growth Span}, defined as the number of years over which GDP per capita growth is computed. The estimated coefficient is negative, but -- in contrast with \cite{de2008meta} -- small and statistically insignificant, indicating limited evidence that longer growth spells systematically shift reported effects in this sample. 

Third, the effect of inequality may vary with countries’ level of development, since the relative importance of transmission channels can differ across income levels \citep{chambers2010relationship, deininger1998new, grundler2018growth}. We include a dummy variable \textit{High-income} equal to one when an estimate is obtained from a sample restricted to high-income economies or is associated to an interaction term with an high-income economy indicator, and zero otherwise. The coefficient in Table~\ref{tab:meta_reg} is positive and statistically significant, suggesting that estimated effects are less negative (or more positive) in advanced-economy settings. This finding aligns with both \citet{de2008meta} and \citet{neves2016meta}.

Fourth, inequality measured net of taxes and transfers may convey different information than inequality measured on market income. Net inequality reflects both market outcomes and the redistributive system and may be more closely related to channels operating through disposable income (e.g., credit constraints and human capital accumulation). By contrast, in the fiscal-policy channel \citep{persson1991inequality}, market inequality may be the relevant object as it affects redistributive pressures and the size of distortionary taxation. To capture this distinction, we include a dummy variable ($Net$ $Inequality$) equal to one when inequality is measured after taxes and transfers and zero otherwise. As shown in Table~\ref{tab:meta_reg}, the coefficient on $Net$ is negative and statistically significant, indicating that net-inequality measures are associated with more adverse estimated growth effects.

Fifth, we control for inequality data quality. Prior to the harmonized dataset introduced by \citet{deininger1996new} and subsequent efforts, studies often relied on ad hoc compiled datasets with limited comparability across countries and time. We include the dummy \textit{High Quality Data}, equal to one when inequality is constructed from survey-based, nationally representative sources with broad income coverage, and zero otherwise.\footnote{High-quality sources include: \citet{deininger1996new}, the Luxembourg Income Study (LIS), the World Income Inequality Database (WIID), the Standardized WIID (SWIID) \citep{solt2016standardized}, the Estimated Household Income Inequality (EHII) by the University of Texas Inequality Project and \citet{iradian2005inequality}.} The coefficient is not statistically significant, consistent with \citet{neves2016meta} but in contrast to \citet{de2008meta}.


Finally, we also test whether the time dimension of the sample influences the results. Studies based on shorter sample periods might capture short-run dynamics or extraordinary events, while longer time horizons help minimize such risks. Therefore, we include \textit{Number of years}, defined as the span between the first and last years of the underlying data. The estimated coefficient is close to zero and statistically insignificant, suggesting that this dimension does not systematically explain variation in reported effect sizes.

\paragraph{Estimator}

Because inequality and growth may be jointly determined, estimators addressing endogeneity -- such as IV or GMM -- may yield different estimates than methods that treat inequality as exogenous. Moreover, since cross-sectional and panel designs are associated with different signs of the effect size, we allow the impact of IV methods to differ by dataset structure. Specifically, we include the interactions $IV \times Cross$ and $IV \times (Panel)$ to distinguish IV estimates in cross-sectional and panel settings respectively. The results suggest that IV methods amplify the negative estimates in cross-sectional designs and the positive estimates in panel designs, although the latter effect is only marginally significant.

We additionally include indicators for estimators used primarily in panel settings. \textit{Difference GMM} captures estimates obtained with difference-GMM methods, and \textit{Fixed effects} identifies fixed-effects estimators. Both coefficients are positive and highly significant, indicating that these estimators are associated with less negative (or more positive) reported effects in panel studies. This result aligns with \citet{de2008meta}, but contrasts with \cite{neves2016meta}.

\paragraph{Covariates} 

Several contributions \citep{persson1991inequality, perotti1996growth} report that the inclusion of regional dummies weakens the effect of inequality on growth, a result also found in both \citet{de2008meta} and \citet{neves2016meta}. To test this effect, we include the variable \textit{Regional Dummy}, which takes the value of 1 if the corresponding estimate includes a regional dummy and 0 otherwise. As with other moderators, we multiply \textit{Regional Dummy} by $Cross-sectional$ since the sign of the effect can vary depending on the structure of the dataset, and most studies that include regional dummies are cross-sectional. Our results confirm that controlling for regional fixed effects reduces the magnitude of the effect size in cross-sectional studies, although it is only marginally statistically significant.

\paragraph{Journal characteristics} 

High quality journal might be more selective in publishing only those papers that conduct rigorous econometric analysis. To control whether the quality of a journal influences the collected effect sizes we add the moderator measuring the journal \textit{Simple Impact Factor} in 2022. The coefficient in Table \ref{tab:meta_reg} is statistically non-significant, suggesting no systematic relationship between journal impact and reported inequality--growth effects in our sample, in line with \citet{neves2016meta}.

\section{Conclusion}
\label{conclusion}

This paper provides a meta-analysis of the empirical literature examining the relationship between income inequality and economic growth. Building on the seminal contributions of \citet{de2008meta} and \citet{neves2016meta}, our analysis extends the existing evidence base along several important dimensions.

First, we update the literature by incorporating studies published over the last decade, thereby capturing recent advances in data availability and econometric practice. Second, unlike previous contributions, we collect all reported estimates from each study rather than relying on author-selected preferred specifications, resulting in a considerably larger dataset of 531 estimates from 33 studies. This approach increases statistical power and reduces the influence of individual studies. Third, in contrast to the previous meta-analysis, we restrict attention to estimates based on the Gini coefficient, ensuring conceptual and quantitative comparability across studies and avoiding the conflation of heterogeneous inequality measures that capture fundamentally different distributional concepts. Fourth, we employ state-of-the-art methods to address publication bias, relying on the RoBMA--PSMA framework, and combine this with a multilevel meta-regression that explicitly accounts for dependence among multiple estimates reported within the same study.

Our results indicate that the average effect of income inequality on economic growth is negative and statistically significant, although economically small. At the same time, we find strong evidence of substantial heterogeneity across studies, consistent with the presence of multiple true effect sizes rather than a single, context-independent inequality--growth relationship. The publication-bias analysis reveals strong evidence of selection based on statistical significance, leading to an over-representation of statistically significant results in the published literature, but no evidence of systematic directional bias.

To explain the observed heterogeneity, we estimate multilevel meta-regressions incorporating a broad set of moderators capturing dataset characteristics, econometric design choices, and journal characteristics. These results show that both real-world features and methodological decisions play a key role in shaping reported estimates. Some of them confirm the conclusion of previous meta-analyses. In particular, the adverse impact of inequality is weaker -- or even reversed -- in high-income economies compared to developing countries. Consistent with \citet{neves2016meta}, we found that methodological choices also matter: cross-sectional studies tend to report more negative estimates, an effect that is attenuated by the inclusion of regional controls. Beyond these similarities, our analysis shows that instrumental-variable, fixed-effects, and difference-GMM estimators are associated with more positive estimates in panel settings. By contrast, growth-horizon length, income-data quality, and journal characteristics do not appear to influence reported results.

A central contribution of this study relies in the finding that the type of inequality measure matters: net income inequality has a more negative effect on growth than market-income inequality. This result suggests that channels operating through disposable income -- such as credit constraints and human-capital accumulation -- may be more relevant than channels operating through market-income inequality alone, such as redistributive fiscal pressures.

\clearpage

\addcontentsline{toc}{chapter}{References}

\bibliography{bibliog}
\appendix

\renewcommand{\thefigure}{A\arabic{figure}}
\setcounter{figure}{0}
\newpage
\section{Appendices}

\subsection{Prisma Diagram}\label{app::prisma}

\begin{figure}[h]
\hspace*{-1.5cm} 
\begin{tikzpicture}[
   node distance=2.5cm and 0.8cm,
  process/.style={rectangle, draw, text width=6cm, align=center, minimum height=1.1cm, font=\small},
  smallbox/.style={rectangle, draw, text width=6cm, align=left, minimum height=1.1cm, font=\small},
  arrow/.style={-{Latex}, thick},
  phase/.style={rotate=90, font=\bfseries\footnotesize}
]

\node[phase] at (-4.4, 2.9) {Identification};
\node[phase] at (-4.2, -1.1) {Screening};
\node[phase] at (-4.2, -5) {Eligibility};
\node[phase] at (-4.2, -8.4) {Included};

\node[process] (db) at (0,3) {3,461 records identified through database searching\\(3,245 on Scopus, 216 on Google Scholar)};
\node[process, right=of db] (snow) {25 additional records identified through backward snowballing};

\node[process, below=of db, xshift=3.4cm] (screened) {3,486 records screened};
\node[process, right=of screened] (excluded1) {3,438 records excluded based on abstract\\};

\node[process, below=of screened, yshift=-4mm] (fulltext)
{48 full-text articles assessed for eligibility};

\node[smallbox, right=of fulltext, yshift=-0mm] (excluded2) {15 full-text articles excluded with reasons:\\
- 2 are working papers\\
- 3 national level\\
- 3 no control for initial GDP\\
- 1 dependent variable not based on GDP\\
- 1 standard error or t-stat not reported\\
- 2 inequality index based on concepts other than income\\
- 3 based on inequality indices other than Gini};

\node[process, below=of fulltext] (included) {33 studies included in the meta-analysis};

\draw[arrow] (db) -- (screened);
\draw[arrow] (snow) -- (screened);
\draw[arrow] (screened) -- (fulltext);
\draw[arrow] (fulltext) -- (included);
\draw[arrow] (screened.east) -- (excluded1.west);
\draw[arrow] (fulltext.east) -- (excluded2.west);

\end{tikzpicture}
\caption{PRISMA Flow Diagram}
\label{fig:prisma}
\end{figure}

\newpage
\subsection{Descriptive Statistics}\label{app::descriptives}

\setcounter{table}{0} 
\renewcommand{\thetable}{A\arabic{table}} 
\begin{table}[htbp]\label{tab::desriptives}
\centering
\caption{Descriptive Statistics}
\begin{tabular}{lccccc}
\hline
\textbf{Variable} & \textbf{Obs} & \textbf{Mean} & \textbf{Std. Dev.} & \textbf{Min} & \textbf{Max} \\
\hline
Median Year             & 531 & 1983.003 &	5.687 &	1972.5 & 	2002 \\
Net Inequality                & 531 & .348 &	.477	0 & 0      & 1    \\
Cross-sectional         & 531 & .277	& .448  & 0      & 1    \\
High-income Country                & 531 & .111	& .315  & 0      & 1    \\
Low-income Country                    & 531 & 0.043 &	.204  & 0      & 1    \\
Growth Span             & 531 & 10.192 &	9.491  & 1      & 40   \\
High Quality Data           & 531 & .893 &	.31  & 0      & 1    \\
Number of Years         & 531 & 34.889 &	11.704	& 8	& 56  \\
OLS (Cross - sectional) & 531 & .207 &	.406  & 0      & 1    \\
OLS (Pooled)            & 531 & .036 &	.186  & 0      & 1    \\
IV (Cross - sectional)                   & 531 & .041	 & .199  & 0      & 1    \\
IV (Panel)                & 531 & .028 &	.166  & 0      & 1    \\
Fixed Effect            & 531 & .143 &	.351  & 0      & 1    \\
Difference GMM                & 531 & .058 &	.235  & 0      & 1    \\
System GMM                 & 531 & .39 &	.488  & 0      & 1    \\
Regional dummies$\times$Cross-sectional     & 531 & .019 &	.136  & 0      & 1    \\
Time Dummy                  & 531 & .548 &	.498  & 0      & 1    \\
Publication Year                & 531 & 2009.437 &	8.698 &	1995	 & 2021 \\
Simple Impact Factor 2022                & 531 & 3.314 &	2.569 &	.4	& 10.7 \\
\hline
\end{tabular}
\label{tab::desriptives}
\end{table}

\newpage
\subsection{Robust Bayesian Meta-Analysis} \label{app:robma}

Let $\theta_{rq}$ denote the observed effect size estimate $q$ from study $r$ with known standard error $\sigma_{rq}$.
All models considered in the RoBMA-PSMA framework share the same sampling distribution
\[
\theta_{rq} \mid \delta_{rq} \sim \mathcal{N}(\delta_{rq}, \sigma_{rq}^2),
\]
and differ with respect to assumptions about the presence of the effect, the between-study heterogeneity and selective publication.
Both fixed-effects models ($\delta_{rq} = \mu$) and random-effects models 
($\delta_{rq} \mid \mu, \tau \sim \mathcal{N}(\mu, \tau^2)$) are included in the model ensemble.
The overall effect $\mu$ is assigned a weakly informative normal prior $\mu \sim \mathcal{N}(0, 2^2)$.

\noindent\textbf{Selection models and weight functions.}
Publication bias is modeled using selection models in which the probability that a study is observed depends on its statistical significance.
Formally, under a given model specification $H_m$, the likelihood contribution of each observed effect size is weighted by a publication weight function
\[
p(\theta_{rq} \mid \delta_{rq}, H_m) \propto f(\theta_{rq} \mid \mu, \tau)\,\omega(p_{rq}),
\]
where $f(\cdot)$ denotes the Gaussian sampling density, $p_{rq}$ is the two-sided $p$-value associated with $\theta_{rq}$, and 
$\omega(p_{rq})$ is a stepwise-constant function defined over intervals of two-sided $p$-values.

In the present analysis, we consider exclusively two-sided selection mechanisms, which capture preferential publication of statistically significant results regardless of the sign of the effect.
Specifically, the model ensemble includes: (i) a two-sided weight function with a single cutpoint at $p = .05$; (ii) a two-sided weight function with cutpoints at $p = .05$ and $p = .10$.

For each weight function, the relative publication probabilities associated with the $J$ $p$-value intervals are denoted by
$\boldsymbol{\omega} = (\omega_1,\dots,\omega_J)$.
These weights are assigned a Dirichlet prior with concentration parameters
$\boldsymbol{\alpha} = (1,\dots,1)$, corresponding to a uniform prior over all admissible weight configurations.

\noindent\textbf{Small-study effects: PET and PEESE.}
In addition to selection models, RoBMA includes regression-based adjustments for small-study effects.
The Precision Effect Test (PET) models a linear dependence between effect sizes and their standard errors
\[
\theta_{rq} = \mu + \beta_{\text{PET}}\, \sigma_{rq} + \varepsilon_{rq},
\]
whereas the Precision Effect Estimate with Standard Error (PEESE) models dependence on the sampling variance,
\[
\theta_{rq} = \mu + \beta_{\text{PEESE}}\, \sigma_{rq}^2 + \varepsilon_{rq},
\qquad
\varepsilon_{rq} \sim \mathcal{N}(0, \sigma_{rq}^2 + \tau^2).
\]
The regression coefficients are assigned weakly informative Cauchy priors centered at zero,
\[
\beta_{\text{PET}} \sim \text{Cauchy}(0,1),
\qquad
\beta_{\text{PEESE}} \sim \text{Cauchy}(0,5),
\]
allowing small-study effects to operate in either direction.

\noindent\textbf{Bayesian model averaging.}
Let $\mathcal{H} = \{H_1,\dots,H_M\}$ denote the full set of candidate models.
Each model $H_m$ specifies a likelihood $p(\boldsymbol{\theta} \mid \boldsymbol{\delta}_m, H_m)$ and a prior distribution
$p(\boldsymbol{\delta}_m \mid H_m)$ over its parameters $\boldsymbol{\delta}_m = (\mu, \tau, \boldsymbol{\omega}, \beta)$.
The posterior distribution of the parameters under model $H_m$ is given by
\[
p(\boldsymbol{\delta}_m \mid H_m, \boldsymbol{\theta})
=
\frac{p(\boldsymbol{\delta}_m \mid H_m)\,p(\boldsymbol{\theta} \mid \boldsymbol{\delta}_m, H_m)}
     {p(\boldsymbol{\theta} \mid H_m)},
\]
where $p(\boldsymbol{\theta} \mid H_m)$ denotes the marginal likelihood of model $H_m$.

The model ensemble combines effect presence ($\mu \neq 0$ vs.\ $\mu = 0$), heterogeneity
($\tau > 0$ vs.\ $\tau = 0$), and alternative publication-bias mechanisms (two two-sided selection models, PET, and PEESE),
resulting in a total of $M = 2 \times 2 \times 5 = 20$ candidate models.
All models are assigned equal prior probabilities $p(H_m) = 1/M$.
Posterior model probabilities are obtained via
\[
p(H_m \mid \boldsymbol{\theta})
=
\frac{p(H_m)\,p(\boldsymbol{\theta} \mid H_m)}{p(\boldsymbol{\theta})},
\]
where $p(\boldsymbol{\theta}) = \sum_{m=1}^M p(H_m)\,p(\boldsymbol{\theta} \mid H_m)$.

Inference on any parameter of interest $\boldsymbol{\delta}$ is obtained through Bayesian model averaging,
\[
p(\boldsymbol{\delta} \mid \boldsymbol{\theta})
=
\sum_{m=1}^M p(\boldsymbol{\delta}_m \mid H_m, \boldsymbol{\theta})\,p(H_m \mid \boldsymbol{\theta}),
\]
which propagates uncertainty about the presence and functional form of publication bias directly into the final posterior inference.

Finally, the Bayes Factors (BF) for a given model component (the presence of the effect, heterogeneity or publication bias) are defined as the ratio of posterior odds to prior odds across the corresponding sets of models,
\[
BF
=
\frac{\sum_{i \in \mathcal{M}_1} p(H_i \mid \boldsymbol{\theta})}
     {\sum_{j \in \mathcal{M}_0} p(H_j \mid \boldsymbol{\theta})}
\Bigg/
\frac{\sum_{i \in \mathcal{M}_1} p(H_i)}
     {\sum_{j \in \mathcal{M}_0} p(H_j)},
\]
where $\mathcal{M}_1$ and $\mathcal{M}_0$ denote, respectively, the sets of models that include or exclude the component of interest.

\newpage
\subsection{Bayesian Model Averaging (BMA) for moderators selection} \label{app:bma}
\begin{table}[htbp]
\centering
\begin{threeparttable}
\centering
\caption{BMA Results for Moderator Screening}
\label{tab:bma_moderators}
\begin{tabular}{lcc}
\hline\hline
Moderator & PIP & Included \\
\hline
Candidates &  &  \\
\hspace{8mm}\textit{Net Inequality} & 1.000 & Yes \\
\hspace{8mm}\textit{Fixed Effect} & 1.000 & Yes \\
\hspace{8mm}\textit{Simple Impact Factor} & 1.000 & Yes \\
\hspace{8mm}\textit{Number of Years} & 0.996 & Yes \\
\hspace{8mm}\textit{Difference GMM} & 0.992 & Yes \\
\hspace{8mm}$IV \times cross$ & 0.978 & Yes \\
\hspace{8mm}\textit{High-income Country} & 0.964 & Yes \\
\hspace{8mm}\textit{Regional Dummy} $\times cross-sectional$ & 0.354 & Yes \\
\hspace{8mm}$IV \times (panel)$ & 0.328 & Yes \\
\hspace{8mm}\textit{Cross} & 0.248 & Yes \\
\hspace{8mm}\textit{Growth Span} & 0.200 & Yes \\
\hspace{8mm}\textit{High Quality Data} & 0.193 & Yes \\
\hspace{8mm}\textit{Time Dummy} & 0.094 & No \\
\hspace{8mm}$OLS \times cross$ & 0.083 & No \\
\hspace{8mm}$OLS \times (panel)$ & 0.060 & No \\
\hspace{8mm}\textit{System GMM} & 0.043 & No \\
\hspace{8mm}\textit{Median Year} & 0.039 & No \\
\hspace{8mm}\textit{Publication Year} & 0.035 & No \\
\hspace{8mm}\textit{Low-income Country} & 0.034 & No \\
\hline
Always included &  &  \\
\hspace{8mm}\textit{Constant} & 1.000 & Yes \\
\hspace{8mm}\textit{Standard Errors} & 1.000 & Yes \\
\hline\hline
\end{tabular}
\begin{tablenotes}
\footnotesize
\item Notes: The table reports Posterior Inclusion Probabilities (PIP) from Bayesian Model Averaging (BMA). 
Candidate moderators were retained for the subsequent multilevel meta-regression if their PIP were above 0.1. 
To account for potential direction-related small-study effects, the standard error of the reported estimate is forced to be included in all BMA specifications.
\end{tablenotes}
\end{threeparttable}
\end{table}

\newpage
\subsection{Other Inequality Indices}\label{app:other_ind}

To compare our results with the most recent meta-analysis on this topic \citep{neves2016meta}, we repeat the meta-regression analys by including also estimates based on inequality indices other than Gini. When the analised index expresses a measure of ``equality" rather than inequality (e.g., share of total income held by the bottom 40\% of the distribution) we premultiply the index by -1 in order to express all measures as inequality indices. Note that this exercise is performed solely to ensure comparability with the existing meta-analysis. We recognize that the indices considered are not quantitatively and conceptually comparable, and therefore the results should be interpreted with appropriate caution.

\subsubsection{Dataset}

The new dataset includes 764 estimates from 36 studies. Similar to Table~\ref{tab::dataset}, Table~\ref{tab::dataset_appendix} displays key characteristics of the papers in the dataset, such as the number of observations (Column 3), whether the analytical framework is cross-sectional (\textit{CS}) or \textit{Panel} (Column 4), the average effect size $\theta$ within each study (Column 5), whether inequality is measured \textit{Gross} or \textit{Net} of taxes and transfers (Column 6), the type of countries included (Column 7), and the estimation method used (Column 8).

\newgeometry{left=1.5cm,right=1.5cm,top=3cm,bottom=3cm}

\begin{table}[h!]
\small
  \caption{Dataset characteristics}
    \begin{threeparttable}
      \resizebox{7.4in}{!}{%
        \begin{tabular}{ | c | l | c | l | c | c | l | l | l }
          \hline
          \hline
          \textbf{N} & \textbf{Study} & \textbf{Observations} & \textbf{Structure} & \textbf{Average $\theta$} &  \textbf{Income Concept} & \textbf{Country type} & \textbf{Estimation Method}\\ 
          \hline
          \hline
1&Alesina and Rodrik (1994)&32&CS&-.182625&Gross, Net&1&OLS \tabularnewline
2&Persson and Tabellini (1994)&21&CS, Panel&-.0276152&Gross&1&OLS,IV,WLS \tabularnewline
3&Clarke (1995)&28&CS&-.0540071&Gross, Net&2&OLS,IV,WLS \tabularnewline
4&Perotti (1996)&12&CS&-.097&Unspecified&1&OLS \tabularnewline
5&Galor and Zang (1997)&25&CS&-.0287075&Gross&1&OLS \tabularnewline
6&Li and Zou (1998)&40&CS, Panel&.087275&Gross, Net&2&OLS,FE,RE \tabularnewline
7&Deininger and Squire (1998)&12&CS&-.0236667&Gross, Net&1&OLS \tabularnewline
8&Tanninen (1999)&9&CS&-.1402656&Gross, Net&2&OLS \tabularnewline
9&Knell (1999)&3&CS&-.0442333&Unspecified&1&OLS  \tabularnewline
10&Barro (2000)&13&Panel&.0006923&Gross&1&RE,IV \tabularnewline
11&Sylwester (2000)&6&CS&-.0531667&Gross, Net&2&OLS,IV \tabularnewline
12&Mo (2000)&20&CS&-.194605&Unspecified&2&IV \tabularnewline
13&Forbes (2000)&51&CS, Panel&.1596078&Unspecified&1&OLS,FE,RE,GMM \tabularnewline
14&Keefer and Knack (2002)&2&CS&-.067&Unspecified&2&OLS \tabularnewline
15&Castelló  and Doménech (2002)&3&CS&.3056667&Unspecified&2&OLS \tabularnewline
16&Banerjee and Duflo (2003)&14&Panel&.1115&Unspecified&2&FE,RE,GMM \tabularnewline
17&Bleaney and Nishiyama (2004)&16&CS&.0161813&Gross&1&OLS \tabularnewline
18&Odedokun and  Round (2004)&8&CS&-.01255&Unspecified&2&OLS \tabularnewline
19&Voitchosky (2005)&37&Panel&.0182719&Net&1&GMM \tabularnewline
20&Knowles (2005)&8&CS&-.0283333&Gross, Net&2&OLS \tabularnewline
21&Sakar (2007)&2&CS&-.0555556&Unspecified&2&OLS \tabularnewline
22&Noh and Yoo (2008)&4&Panel&.276525&Unspecified&2&FE \tabularnewline
23&Huang, Lin and Yeh (2009)&4&Panel&.061625&Unspecified&1&OLS,SEM-GMM \tabularnewline
24&Castellò-Climent (2010)&96&Panel&-.0014167&Unspecified&1&GMM \tabularnewline
25&Chambers and Krause (2010)&3&Panel&.0024667&Gross, Net&2&GMM \tabularnewline
26&Woo (2011)&18&CS&-.0673889&Unspecified&2&OLS \tabularnewline
27&Davis and Hopkins (2011)&25&CS, Panel&-.0059299&Unspecified&2&OLS,FE,RE,BE,SEM-OLS \tabularnewline
28&Herzer and Vollmer (2012)&6&Panel&-.0123333&Gross&1&ECM \tabularnewline
29&Halter et al. (2014)&32&Panel&-.0062125&Gross&2&GMM \tabularnewline
30&Bartak and Jabłoński (2018)&45&Panel&-.1426711&Gross, Net&3&GMM \tabularnewline
31&Brueckner and Lederman (2018)&6&Panel&.2333333&Unspecified&1&FE,IV \tabularnewline
32&Grundler and Scheuermayer (2018)&34&Panel&-.1573729&Net&2&GMM \tabularnewline
33&Berg et al. (2018)&22&Panel&-.1523227&Net&2&GMM \tabularnewline
34&El-Shagi and Shao (2019)&2&Panel&.6985&Unspecified&2&FE \tabularnewline
35&Scholl and Klasen (2019)&8&Panel&-.0749&Net&2&FE,IV,GMM \tabularnewline
36&Marrero and Serven (2021)&97&Panel&-.0299071&Net&1&OLS,FE,GMM \tabularnewline
        \hline
        \end{tabular}%
      }
\begin{tablenotes}
    \scriptsize
    \item \begin{minipage}[t]{1\textwidth}
    Notes: In Column 7 (Country type): (1) refers to the inclusion of both high-income and least developed countries, controlling for the level of development; 
    (2) to the inclusion of both type of countries without controlling for the level of development; 
    (3) to the inclusion of only high-income countries. 
    In Column 8 (Estimation Method): WLS stands for Weighted Least Squares; 
    FE for Fixed Effects; RE for Random Effects; BE for Between Estimation; ECM for Error Correction Model.
    \end{minipage}
\end{tablenotes}
    \end{threeparttable}
  \label{tab::dataset_appendix}
\end{table}

\newgeometry{left=3cm,right=3cm,top=3cm,bottom=3cm}

\subsubsection{Meta-Regression Analysis}

\begin{table}[h!]
\centering
\caption{Multilevel Meta-Regression}
\label{tab:meta_reg_appendix}
\begin{threeparttable}
\footnotesize
\begin{tabular}{lc}
\toprule
Parameters & Coefficients \\
\midrule

\hspace{8mm}\textit{Standard error} 
& $-0.5837^{***}$ \\
& $(0.0991)$ \\

\midrule

\textbf{Dataset characteristics} & \\

\hspace{8mm}\textit{Cross-sectional} 
& $-0.1017^{***}$ \\
& $(0.0165)$ \\

\hspace{8mm}\textit{Growth Span} 
& $0.00039$ \\
& $(0.00066)$ \\

\hspace{8mm}\textit{High-income} 
& $0.0160^{**}$ \\
& $(0.0068)$ \\

\hspace{8mm}\textit{Net Inequality} 
& $-0.0425^{*}$ \\
& $(0.0229)$ \\

\hspace{8mm}\textit{High-quality data} 
& $-0.0102$ \\
& $(0.0300)$ \\

\hspace{8mm}\textit{Number of years} 
& $-0.00107^{***}$ \\
& $(0.00022)$ \\

\midrule
\textbf{Estimator} & \\

\hspace{8mm}\textit{IV $\times$ cross} 
& $-0.00612$ \\
& $(0.0151)$ \\

\hspace{8mm}\textit{IV $\times$ (panel)} 
& $0.0962^{**}$ \\
& $(0.0460)$ \\

\hspace{8mm}\textit{Difference GMM} 
& $0.1425^{***}$ \\
& $(0.0135)$ \\

\hspace{8mm}\textit{Fixed Effects} 
& $0.1250^{***}$ \\
& $(0.0119)$ \\

\midrule
\textbf{Covariates} & \\

\hspace{8mm}\textit{Regional dummy $\times$ cross} 
& $0.0352^{**}$ \\
& $(0.0171)$ \\

\midrule
\textbf{Journal characteristics} & \\

\hspace{8mm}\textit{Simple Impact Factor (2022)} 
& $0.00380$ \\
& $(0.00429)$ \\

\midrule
Observations & 764 \\
Studies & 36 \\
\bottomrule
\end{tabular}

\begin{tablenotes}
\item \textit{Notes:} Standard errors in parentheses. 
$^{*} p<0.10$, $^{**} p<0.05$, $^{***} p<0.01$. 
Results are based on a multilevel REML meta-regression with random intercepts at the study and estimate levels.
\end{tablenotes}
\end{threeparttable}
\end{table}


\end{document}